# The X-Files: Investigating Alien Performance in a Thin-client World

Dr. Neil J. Gunther

Performance Dynamics Company™
Castro Valley, CA 94552, U.S.A.
www.perfdynamics.com

**Abstract**

Many **scientific applications** use the **X11** window environment-—an **open source** windows GUI standard employing a client/server architecture. X11 promotes: distributed computing, thin-client functionality, cheap desktop displays, compatibility with heterogeneous servers, remote services and administration, and greater maturity than newer web technologies. This paper details the author's investigations into close encounters with alien performance in X11-based seismic applications running on a **200**-node **cluster**, backed by **2 TBytes** of mass storage. End-users cited two significant **UFOs** (Unidentified Faulty Operations) a) long application **launch times** and b) poor interactive **response times**. The paper is divided into three major sections describing Close Encounters of the **1st Kind**: citings of UFO experiences, the **2nd Kind**: recording evidence of a UFO, and the **3rd Kind**: contact and analysis. UFOs do exist and this investigation presents a valuable test case against which to weigh the claims of vendors when procuring performance management and analysis tools.

## 1   Introduction[†]

This paper describes techniques used to uncover the true nature of some performance gremlins or aliens inhabiting UFO's (Unidentified Faulty Operations) in a large-scale client/server computing environment.

Despite the usual hype by vendors, no single performance management tool could have unmasked these aliens. As the reader will soon see, it took a number different strategies and tools (some ad hoc) to identify these UFOs. In that sense this investigation presents a valuable test case against which to weigh vendor claims when procuring performance management tools for large-scale distributed computer systems.

## 2   The Production Environment

This investigation took place during a consulting engagement and the client has requested that all names and private product identifications be changed to maintain confidentiality. I have duly met that request. In any performance analysis project it is important to gain a good understanding of the computing environment, both the hardware platforms, networks, and the software applications running on those platforms.

### 2.1   Production Applications

The production environment supported a large number of geophysicists doing analysis of raw seismic data collected from either land or marine based explorations. To facilitate the geophysical analysis, the raw seismic data is transformed both numerically and graphically. Once the raw data files have been transferred to the appropriate computational facility, the applications become mostly CPU-intensive. Some applications are run interactively and others are run in batch mode. In this paper we concentrate on the performance of three interactive applications hereafter denoted: applicationA, applicationB, and applicationC. Next, we review the server and networking hardware.

---

[†] © 1999, 2000 Performance Dynamics. All Rights Reserved. No part of this document may be reproduced, in any form, without the prior written permission of the author. Permission is granted to Hiper99, to publish this article in the 1999 confernece proceedings.

**2.2 Cluster and Networking Architecture**

The central platform was a 4 by 50-node [IBM] SP2 cluster running AIX 4.1 [CEV96] as the operating system. These clusters formed the Seismic Processing System (SPS). Seismic data coming form the field is stored on tapes. On request, tapes are loaded in robotic silos and staged onto DASD mass storage on an IBM 390 mainframe. The mainframe acts as the bridging server between the silos and the SPS platform, providing data retrieval across 2TB of mass storage.

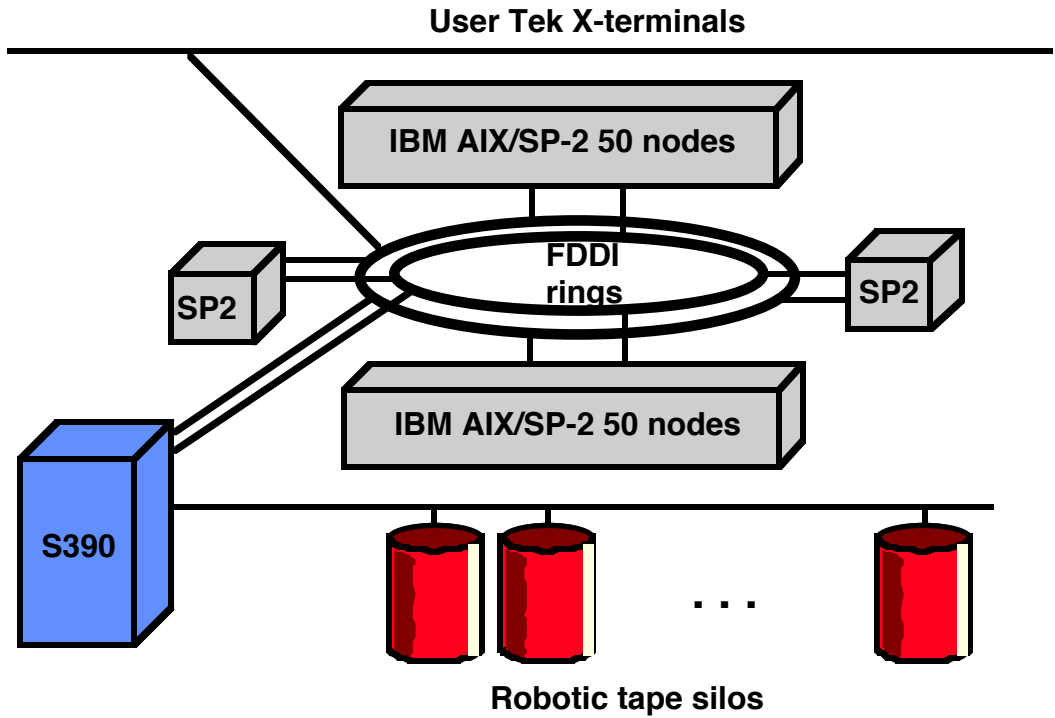

Fig 1:   Schematic illustration of the production seismic data processing environment.

Even though the system was not an HP platform, our findings (and perhaps more importantly, our **methods**) should be applicable to investigating alien X11 performance on any platform.

## 3   The X11 Windows Architecture

Many commercial and scientific UNIX applications use the X11 standard window environment [XCON].

**3.1   X11 Virtues**

X11 windows is based on a client/server software architecture that promotes and offers:
- Distributed computing
- Thin-client functionality
- Compatibility with heterogeneous servers
- Remote services and management: files, fonts, database, etc.

- Greater maturity than newer Java-based technologies.

The client/server terminology used for X11 is reversed from the usual conventions. As shown schematically in Fig. 2, the Xclient does the computational work and tells the Xserver screen what to draw. There is a separate X11 messaging protocol [DAN94] supporting communication between the Xclient and the Xserver.

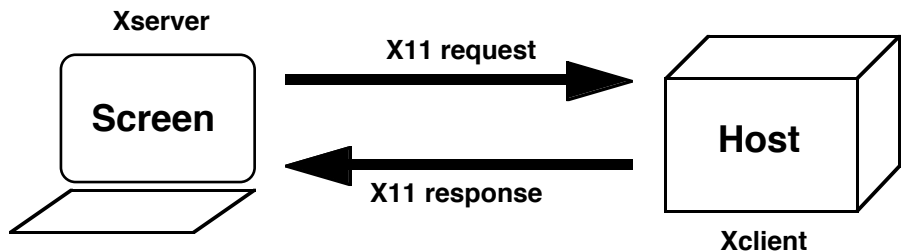

Fig 2: Relationship between client and sever in the X11 architecture.

Rather than take up space unnecessarily here, we refer the interested reader to the plentiful literature on X11 windows and its architecture [LEE98].

### 3.2 Thin-client Trends

Although X11 is arguably a 1980's technology, the X-based concepts of remote font servers, distributed computing cycles and cheap desktops are looking attractive once again.
In the 90's, there has been a dramatic rush to place cheap PCs on every user's desktop thus putting more CPU-horsepower at the disposal of each user and alleviating the kind of contention seen on the time-share systems of the 80's. But it seems there is nothing new in computer technology.
A current trend in corporate computing is to avoid the canonical 2-year cyclical expense of upgrading PCs by reverting to thin-clients (dumb window desktops) and putting the computational horsepower back into heterogeneous servers instead of the desktop. These modernized dumb desktops come with a windows GUI that can support any of X-windows (e.g., NCD, Tek), MS Windows (e.g., WinTerm), or Mac, and the servers can also be heterogeneous (Mainframes, UNIX[NCD], NT[CIT]).
The issue that is still often overlooked, however, is the potential performance impacts that are not even common knowledge amongst X aficionados [YOU94].

## 4 Encounters of the 1st Kind – Citings

As with all scientific investigations of UFO citings, one has to begin with rumors and verbal reports and one has to attempt to retain most of this information for some time until serious leads begin to reveal themselves.

### 4.1 The Rumors

It is very important to talk with **people** when trying to understand UFOs. This the most efficient way to quickly get a sense of the observed phenomena, its variety, its scale, and frequency of occurrence.
Similarly, everyone has their pet theories to explain the strange effects they have observed. In this case, the favorite suspects included:
- FDDI network
- NFS file system
- SP2 interconnect

In such a large-scale system there are many pieces that people can guess at as the culprit. Our job was to listen but verify.

### 4.2 Juggling Facts

As well as the potential technical folklore, there is always a certain amount of political finger-pointing going on. Your urge is to want to discount most of these opinions but you really can't. Something said might be useful.

This investigator tends to juggle many of these opinions and reports in the back of his mind just in case one of them provides a clue or hint when all other roads seem blocked. For this particular investigation, all of these turned out to be false leads.

## 5 Encounters of the 2nd Kind – Evidence

From these discussions with witnesses, it became clear that there were two characteristic time scales reflective of UFO application performance:
- Application **Launch** Times - the time it takes to load the executable and otherwise establish the first instance of an application in memory.
- Interactive **Response** Times - the time for the application to respond to user actions (e.g., typing, mousing), once it has been launched.

For each of these time scales, we gathered data to determine both the **mean** (arithmetic average) of those times and the statistical **variance** about that mean. In this way, we were able to pass from purely **qualitative** statements about "erratic responsiveness" to a sound basis for doing **quantitative** performance analysis.

To gather statistically meaningful evidence it is important to:
- **Measure** times with appropriate accuracy
- Repeat measurements on a **scheduled** basis
- **Sample** long, contiguous periods of time
- **Record** these data in a permanent log

One way to meet these criteria is to build a **benchmark** that performs the same actions as a typical production user.

### 5.1 Injection Benchmarking

The author built the first such benchmark as a single UNIX shell-script that launched all three applications in succession, measured their respective launch times on the production system every 15 minutes on a 7 x 24 basis, and logged the results for each 15 minute cycle.

This scenario simulated the actions of a **single user** interacting with the production environment in the presence of both *real* users, and other (undetermined) workloads. Running the benchmark provided the first statistically meaningful performance database by which to compare user citings.

Figures 3 and 4 show plots of typical benchmark data recorded over a 3 hour period from 5pm to 8pm. Fig. 3 is a **histogram** that shows the **raw data** as it was really measured—a set of **discrete** points in time. Fig. 4 shows the raw data in a less cluttered visual form—a set of three **continuous** lines. The latter form is more appropriate for **time-series** analysis. Note the large spike in the AppC launch time near timestamp 18:22. At this time of the day, it is likely that some **serial batch** workloads may have been running on the production system. Such raw data alone is not sufficient for performance analysis but it does give a visual impression that a high degree of variance is present in the application launch times.

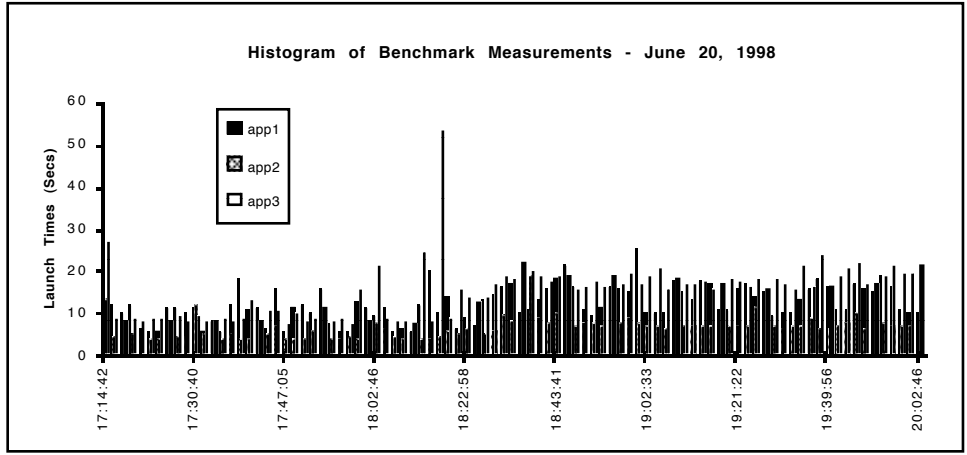

Fig 3:    Sample raw injection benchmark data.

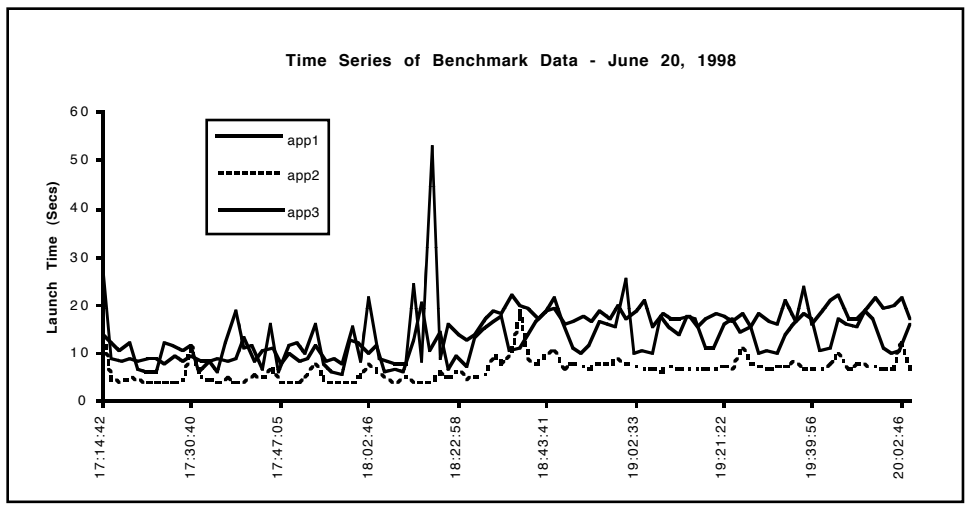

Fig 4:    Data from Figure 3 represented as a time series plot for easier visualization.

To record launch times the injection benchmark required only relatively straightforward X-window intervention and control, since each of the three applications was launched successively so there was only one X-window to monitor and control at each time.

**5.2   Multi-user Scalability Tests**

A separate set of tests on an isolated SUT were necessary to measure and assess the scalability [GUN96] of multiple users on the same sp2node but **isolated** from the presence of other users, and other (undetermined) workloads.  The purpose here was to determine the level of computing resource contention due specifically to application workloads without the presence of other background workloads. For technical reasons, we were only able to use AppA for these tests.

To drive and control multiple instances of AppA , it was necessary to use a special tool for this purpose. LoadRunner® [Mer98] was the tool selected. We had expected to drive the isolated SP2 SUT from another SP2 driver-node while controlling the tests from a standard Tektronix X-terminal. The author finally determined that this was not possible due to limitations within the LoadRunner tool architecture. The final test configuration, therefore, involved a Windows-NT workstation to act as the test driver.  Although this was not a totally faithful representation of the standard SPS environment, it did allow us to gather meaningful performance data up to the point where the SP2 SUT reached a thrashing state. The driver and SUT configurations are summarized in Table 1.

| System | Node | CPU | MHz | RAM | VM |
|--------|--------|----------|-----|-------|--------|
| SUT | sp2node | Power2 | 120 | 512MB | 2048MB |
| DRV | DeskPro | Pentium II | 233 | 98MB | 186MB |

Tab 1: Driver and SUT system configurations.

The measurement of AppA interactive **response times** was much more complicated than the measurement of its launch times. It required recognizing sibling X-window contexts, controlling and tracking mouse movements within those contexts, and exiting sibling windows correctly and in the right order. This level of control demanded a tool that can capture the relevant X activity *once* and generate a script for an indefinite number of later replays. Several products are available on the market that claim to do this. Once again, LoadRunner came closest to meeting all these requirements, although it did not and could not operate in our standard production environment.

During the course of this engagement, several variants of the original benchmark script were redesigned to match the increasing complexity of other supporting measurement tools such as Baseline® User Probes [TQC98]. Another requirement was to recast benchmark into LoadRunner's TSL scripting language[Mer98].

## 6  Encounters of the 3rd Kind – Contact

Using the first of the benchmark variants described in section 5, we were able to determine the performance characteristics of all applications in the launch phase. To see this more clearly, the raw data collected by the Benchmark was analyzed for statistical parameters. This was accomplished by first placing the measured launch times into 3 second buckets and plotting the frequency with which the measurements fall into these buckets (see Fig. 5).

This method provides a simple view of the statistical distribution of the data and from it we were able to determine such statistical parameters as: the mean, standard deviation, sample variance, and sample covariance.

### 6.1 Response Time Distributions

The author wrote a shell-script (*vfstat*) to do statistical analysis on the large amount of collected Benchmark data. A sample of vfstat output for the three measured Seismic applications is shown in Fig. 5 (rotate the page counter-clockwise to view the histogram distribution). Several key characteristics are apparent.

For **applicationA** the measured launch time characteristics can be summarized as:
- Minimum times of about 10 sec.
- Mean times (R) of about 20 sec.
- Standard deviation (SD) of about 30 sec.
- Maximum times of several hundred sec.

For **applicationB** the measured launch time characteristics can be summarized as:
- Minimum times of about 6 sec.
- Mean times (R) of about 10 sec.
- Standard deviation (SD) of about 3 sec.
- Maximum times of 20 sec.

For **applicationC** the measured launch time characteristics can be summarized as:
- Minimum times of about 6 sec.
- Mean times (R) of about 15 sec.
- Standard deviation (SD) of about 5 sec.
- Maximum times of 50 sec.

In all cases, the **asymmetric** distribution with a long tail, is clearly evident. Physically, this means that any application user launching an application would generally see a mean launch time in the range 10-20 seconds, but occasionally would see launch times more than ten times that long. Although launching an application is typically only performed once per session, such a huge variation in launch times can already give the user the unnecessary perception of "erratic" or "poor" application performance.

```
AppA Stats for 728 samples in "bench.log.x"
================================================================
Minm:    5.69     Mean:   14.26    Maxm: 773.94     MDev:    4.35
Var:   811.65    SDev:   28.49    COV:     2.00
GamA:    0.25    GamB:   56.92
Secs |    5%   10%   15%   20%   25%   30%   35%   40%   45%   50%
-----|----+----+----+----+----+----+----+----+----+----+---------
 < 3 |
 3- 6|*
 6- 9|****************
 9-12|********************
12-15|***************************
15-18|*********************
18-21|****
21-24|**
24-27|*
27-30|
30-33|
33-36|
36-39|
39-42|
42-45|
45-48|
48-51|
51-54|
54-57|
```

Fig 5: vfstat output for analysis of injection benchmark data.

During the launch phase, the application was measured to be mostly in a **wait state** on the corresponding SP2 node (i.e., it was neither CPU-bound nor I/O-bound). To resolve further how time was being spent during the launch phase, it was necessary to monitor the actual X-message traffic.

**6.2 Font Service Latencies**

A tool called *Xscope* (provided with the X11R5 source distribution) was used for this purpose. Xscope uses the simple concept of interposing itself between the Xserver (Tek terminal) and the Xclient (SP2 node) in such a way that the Tek thinks it is responding directly to the SP2 and the SP2 thinks it is talking to the Tek. The ambushed X-traffic is decoded on the fly and logged to a trace file. No promiscuous network sniffing is required.

The following two UNIX commands tell Xscope (running on sp2node46) to record X-traffic from Tek terminal htx121 and log it to the file xscope.trace.

```
xscope -htx121 -i0 -v1 >xscope.trace &

appA -display sp2node46:0 1>appA.trace 2>/dev/null &
```

A surprising performance feature is that Xscope imposes no more than about 2% to X-traffic latency. From Xscope traces it was determined that most of the launch time was spent resolving application font requests across the two X-font servers (fs001 and fs002).

AppA, for example, requires about 40 fonts to launch but there was an aggregate of more than 15,000 fonts available. The 40 fonts that best fit the X-request must be searched amongst the thousands of fonts available.

During these experiments, several font protocols were measured for their relative latencies. As well as the two font servers just mentioned (that use the X-protocol to resolve font requests), the Tektronix boot PROM is another source (13 fonts), as well as the Tek boot file system (96 fonts), and NFS-mounted files on Zeus1, Zeus2 and Kepler. The actual number of fonts is shown below:

```
total fonts in Path tcp/fs001.acme.com:7000
   8371
tcp/fs002.acme.com:7000
   7113
/xterms/teknc305/boot/
     96
resident/
     13
tcp/zeus.acme.com:7100
   4781
tcp/kepler.acme.com:7100
   6237
```

The author constructed a simple performance model to confirm that the mean launch time (R) exhibits logarithmic scaling with the **number of font files** (F) on the font servers in the following way. For any two font sources (e.g., **a** and **b**), having a number font files $F_a$ and $F_b$ respectively, the font-query time for path-a can be expressed relative to the query time for path-b as:

$$R_a = R_b \times \frac{\log F_a}{\log F_b}$$

In other words, 1000 fonts will incur a mean launch time 3/2 times longer than if there were just 100 fonts (Fig. 6). Some typical measured times are compared with estimates from the logarithmic model in Table 2.

| Server | Fonts | Type | |
|---|---|---|---|
| **Nfs1** | 8371 | X font | server |
| **Nfs2** | 7113 | X font | server |
| **Nz1** | 4781 | NFS | mounted |
| **Nz2** | 109 | NFS | mounted |

| Times | Measure | Model | %Error |
|---|---|---|---|
| **Rz2** | 9.41 | 9.41 | N/A |
| **Rz1** | 17.01 | 16.99 | -0.09 |
| **Rfs1** | 18.57 | 18.12 | -2.50 |
| **Rfs2** | 16.72 | 17.79 | 6.02 |

Tab 2: Comparison of the logarithm model and data.

Here, the subscript notation fs1 and fs2 refers to the standard font servers (older RS6000 workstations) identified by the network names fs001 and fs002, respectively. The alternative font paths with subscripts z1 and z2 refer to the NFS-mounted servers identified respectively as Zeus1 and Zeus2. Service from Zeus2 gave the best typical launch time ($R_{z2}$) for Seisflow and was therefore used as the basis for normalization. The *minus* sign in the percentage error column indicates that the model *underestimated* the measured time.

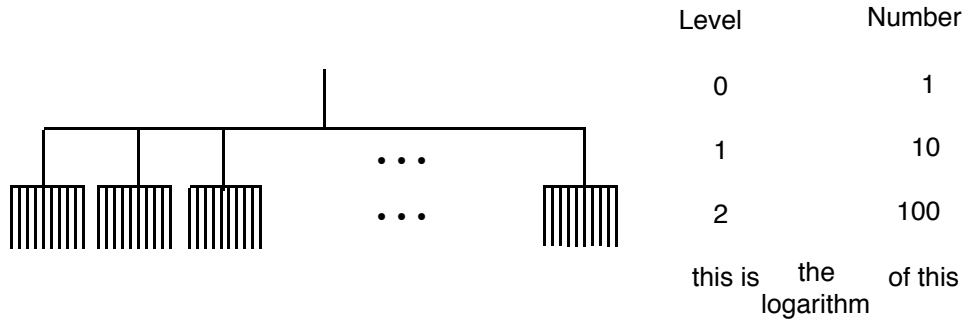

Fig 6: To get a log, you need a tree!

The great value of this simple model is that it immediately suggests the quick and cheap fix of simply paring back the number of fonts on the font servers. Given the complexity of the various subsystems needed to support the operation of X-windows (e.g., Xclient software port, memory accesses, buffering, network load, etc.) it is surprising that none of these things plays a more significant role in determining the mean launch time. As suggested by the log-model, the most important thing is the number of fonts on the font servers. Further experiments with a reduced number of fonts bore this conclusion out (with an error margin of less than ±10%).

**6.3 Application Scalability Tests**

A set of experiments was performed with LoadRunner to measure the increase in launch times as the SP2 node was placed under increasing contention due to increasing user-load.

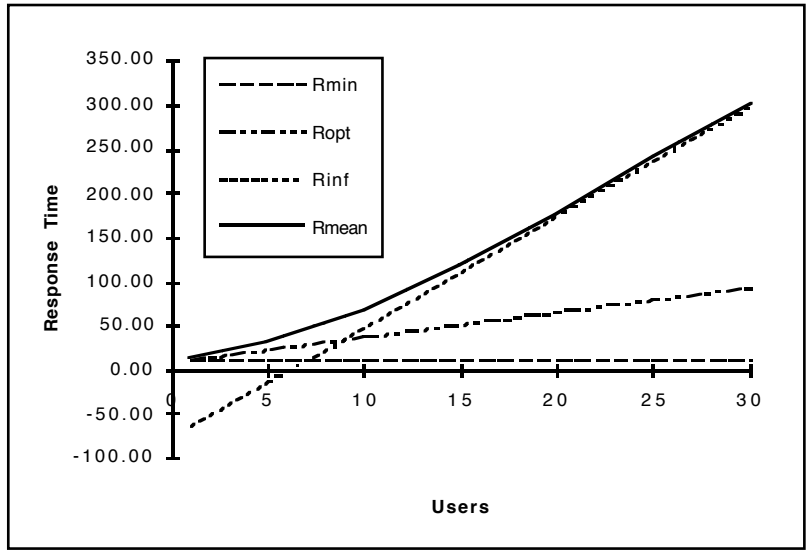

Fig 7: Theoretical mean response curves and asymptotes.

The expected theoretical form of a typical response time loading curve looks like the heavy dark line shown in Fig. 7. The heavy line is bounded by three asymptotes:
- The **minimum** response time (no contention present) labeled **Rmin**
- The **optimistic** bound given that some contention will be present and labeled **Ropt**

- The **asymptotic** rise in response time as the number of users increases and assuming no other limit is reached in the meantime. Labeled **Rinf**.

A complete discussion of the theory underlying this response time curve analysis would take us too far afield. The interested reader can read a more detailed account in [Gun98].

The actual measurements are plotted in Fig. 8 and correspond to the response time histograms described in section 5. Overall, response time data would appear to have the correct characteristic.

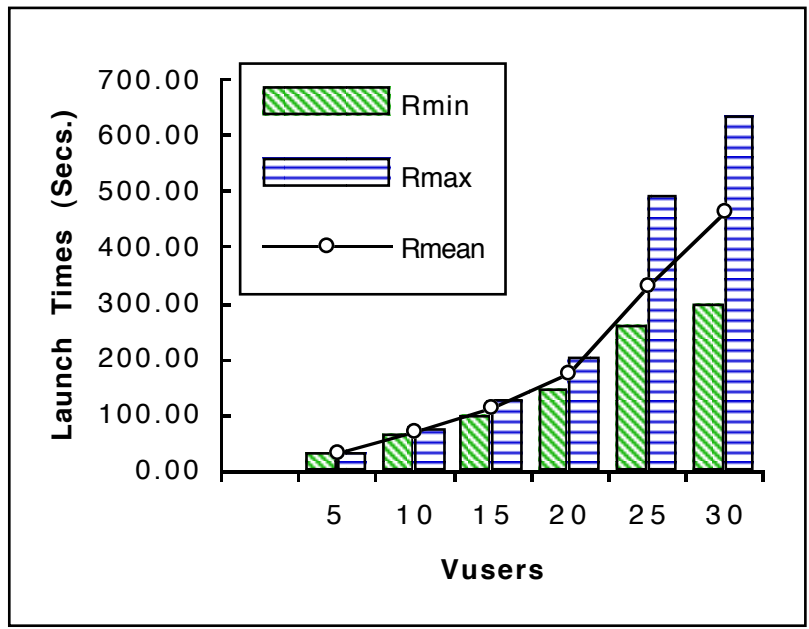

Fig 8: AppA launch times with the mean represented by a line for emphasis.

Closer inspection, however, reveals significant deviations from the expected characteristic of Fig. 7. In Fig. 8 we see that the above 20 LoadRunner Vusers, the mean Seisflow launch times escalate dramatically above the theoretical asymptote (note that the optimistic bound has been elided for clarity since it doesn't offer any new information for our investigations). This can only happen when a secondary effect (such as virtual memory paging) begins to dominate the measured response times. That paging is indeed the correct secondary effect was revealed in *vmstat* output (sampled at 3 second intervals) where it was clear that few free pages were available during the LoadRunner tests on sp2node48.
Optimal loading is determined by the point where the asymptote intersects the lower bound i.e., Vusers = 7. Recalling that these curves are based on *mean* response time values, this choice allows for the inevitable fluctuations (variance) around the optimum while trying to minimize their impact on the Vuser response.
Maximal loading is determined by the point where the response curve rises faster than the theoretical asymptote i.e., Vusers = 20. Fluctuations above this point are likely to occur often and only recover in times that are relatively long compared with those about the optimum.

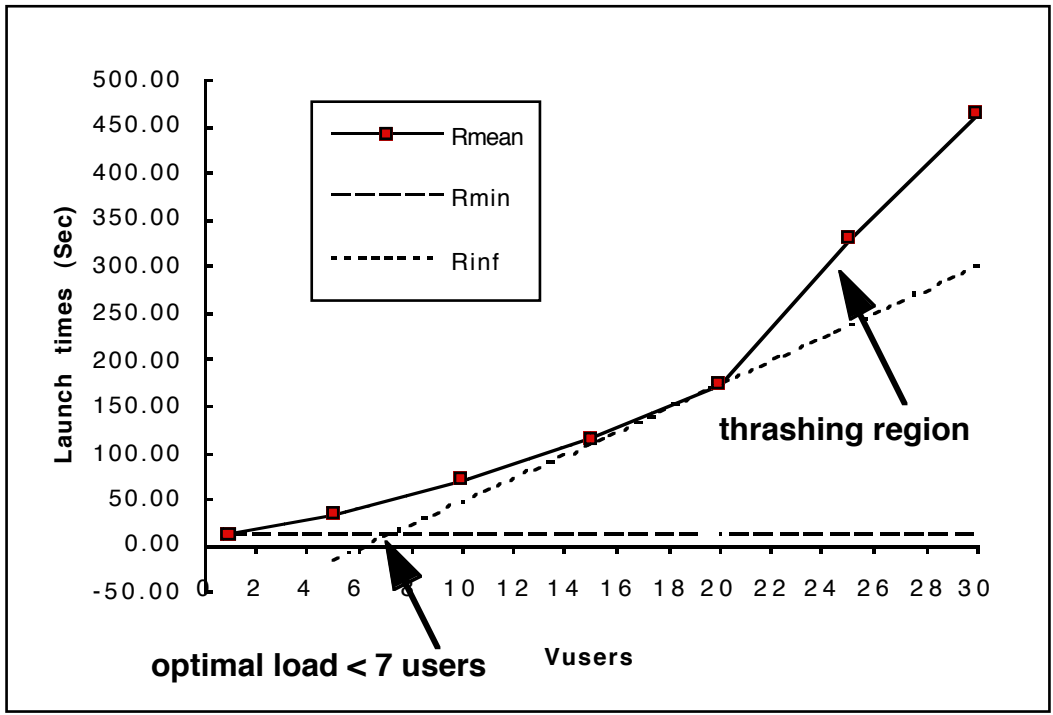

Fig 9: Correct asymptote for the measured mean response curve.

This interpretation is further corroborated by measurements of AppA **interactive** response times. At 20 Vusers the response times for the **interactive** AppA operations of **opening** and **saving** a file, suddenly escalate by an order of magnitude over the times measured for 15 Vusers.

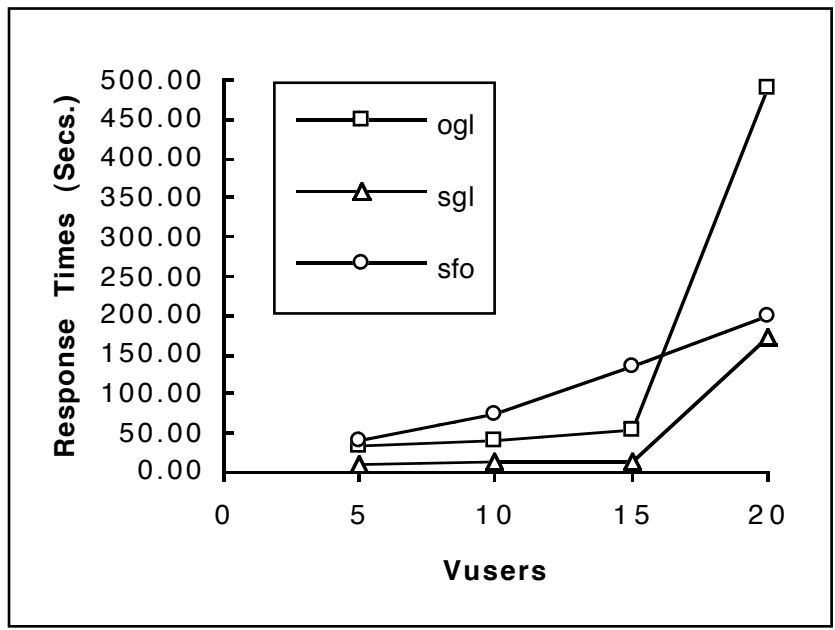

Fig 10: AppA interactive response times where the interactive operations are denoted: sfo = open the AppA application, ogl = open an AppA file, and sgl = save that file.

We did not have time to investigate whether this is also a side-effect of virtual memory paging or some other phenomenon.

**6.4 Network Traffic**

The next few pages show plots of network data reported by LoadRunner [Mer98] at 1 second sampling during the course of running the scalability tests for 1, 15, and 30 Vusers. Clearly the packet intensity is seen to increase as the Vuser load increases but what are these packets? LoadRunner does not tell us, so we turned to the Baseline database. The only significant packet activity reported by Baseline [TQC98] (at a 1 minute sampling rate) was for **NFS packets** with the profile shown in Table 3.

| NFSop | Calls/Sec |
|---|---|
| lookup | 20.00 |
| getattr | 11.70 |
| fsstat | 3.50 |
| commit | 3.30 |
| read | 0.00 |
| write | 0.00 |

Tab 3: Measured throughput of application NFS operations.

The most striking thing about these data, the lack of significant NFSread and NFSwrite operations. The remainder of the NFS operations are related to comparing file names, properties and checking status without the actual transfer of data.

The call rates are also small when compared with the relative call rates supported by various networks:
- Ethernet (10 Mb/sec) supports up to 300 NFS calls/second
- FDDI (100 Mb/sec) supports up to 3000 NFS calls/second
- CSS (400 Mb/sec) should support up to 12,000 NFS calls/second

From this standpoint, we would argue that we are justified in declaring sp2node48 as an isolated node that exhibits very little shared activity or bandwidth contention during the LoadRunner tests (although some file-caching may have taken place in early warm-up runs).

## Conclusion

UFOs do exist! However, we found no evidence that they existed in the rumored locations outlined in section 4 (i.e., the FDDI network, NFS, or cluster switching fabric). The two biggest contributors to poor performance were: distributed font services, and memory and CPU scalability [GUN93] on the SP2 nodes. Both of these performance problems being related to the X11 architecture. The first of these problems has a quick and cheap fix, the second can be addressed with a more effort and expense.

Although it was helpful to have some performance management software running on the SPS platform, no single performance management application that we had available to us, could have enabled us (on its own) to have a close encounter of the 3rd kind. Other commercial performance measurement applications (that were not worth reproducing for ourselves in the time available) could not be adapted faithfully to the SPS environment. Many ad hoc tools and experiments were needed to resolve problems on a system of this scale.